\documentclass[preprint]{revtex4}   
\usepackage{amsfonts}
\usepackage{amsmath}
\usepackage{cancel}
\usepackage{graphicx}
\usepackage{mathcomp}
\usepackage{color}
\usepackage{ulem}
\usepackage{mathtools}
\usepackage{caption}
\usepackage[english]{babel}

\usepackage[utf8]{inputenc}

\allowdisplaybreaks

\newif\iffigures
\figurestrue

\renewcommand{\vec}{\mathbf}

\def\undertilde#1{\mathord{\vtop{\ialign{##\crcr
$\hfil\displaystyle{#1}\hfil$\crcr\noalign{\kern1.5pt\nointerlineskip}
$\hfil\widetilde{}\hfil$\crcr\noalign{\kern1.5pt}}}}}

\begin{document}
\title{Modelling of a miniature microwave driven nitrogen plasma jet  \\ and comparison to measurements}

\author{Michael Klute$^{1}$}
\author{Efe Kemaneci$^{1}$}
\author{Horia-Eugen Porteanu$^{2}$}
\author{Ilija Stefanovi\'{c}$^{3}$}
\author{Wolfgang Heinrich$^{2}$}
\author{Peter Awakowicz$^{4}$}
\author{Ralf Peter Brinkmann$^{1}$}

\affiliation{$^{1}$Theoretical Electrical Engineering, Ruhr University Bochum, Germany}
\affiliation{$^{2}$Microwave Department, Ferdinand-Braun-Institut, Berlin, Germany}
\affiliation{$^{3}$Institute of Technical Sciences, Serbian Academy of Sciences and Arts, Belgrade, Serbia}
\affiliation{$^{4}$Electrical Engineering and Plasma Technology, Ruhr University Bochum, Germany}
\date{\today}

\begin{abstract}
The MMWICP (Miniature MicroWave ICP) is a new plasma source using the induction principle. Recently Klute et al. presented a mathematical model for the electromagnetic fields and power balance of the new device. In this work the electromagnetic model is coupled with a global chemistry model for nitrogen, based on the chemical reaction set of Thorsteinsson and Gudmundsson and customized for the geometry of the MMWICP. The combined model delivers a quantitative description for a non-thermal plasma at a pressure of $p=1000\,\mathrm{Pa}$ and a gas temperature of $T_\mathrm{g}=650\mbox{-}1600\,\mathrm{K}$. Comparison with published experimental data shows a good agreement for the volume averaged plasma parameters at high power, for the spatial distribution of the discharge and for the microwave measurements. Furthermore, the balance of capacitive and inductive \linebreak coupling in the absorbed power is analyzed. This leads to the interpretation of the discharge regime at a electron density of $n_\mathrm{e} \approx 6.4 \!\times\!10^{18} \, \mathrm{m}^{-3}$ as $E/H$-hybridmode with an capacitive and inductive component.
\end{abstract}

\maketitle

\pagebreak

%%%%%%%%%%%%%%%%%%%%%%%%%%%%%%%%%%%%%%%%%%%%%%%%%%%%%%%%%%%%%%%%%%%%%%%

\section{Introduction}

The MMWICP is a promising new plasma source which transfers the principle of inductive coupling successfully to a small jet and was first described in \cite{ref1}. It is based on a specially designed resonator that acts as a LC-circuit with a high quality factor $Q$. During experimental operation, the MMWICP was characterized using optical emission spectroscopy (OES), optical imaging and \grq Hot-S-Parameter' spectroscopy \cite{ref2,ref3}. The principle of inductive coupling was proven. Among the results, the high electron density of up to $n_\mathrm{e} \approx 3.5 \!\times\!10^{19} \, \mathrm{m}^{-3}$ is particularly remarkable. The measured gas temperature of $T_\mathrm{g}=650\mbox{-}1600\,\mathrm{K}$ shows that the plasma is far away from thermal equilibrium. The new source is therefore a potential tool for various technical applications, such as plasma surface treatment \cite{application1}, gas conversion \cite{application2}, material processing \cite{application3}, analytic spectroscopy \cite{application4} etc. Although the operation was successful for a wide pressure range of $p =50\mbox{-}1000\,\mathrm{Pa}$ and with different gases (argon, nitrogen,\linebreak oxygen), the source is best studied for a nitrogen plasma at $p =1000\,\mathrm{Pa}$. Nitrogen allows a very sensitive determination of the electron density using OES and the pressure of $p =1000\,\mathrm{Pa}$ is high enough to describe the plasma kinetics in a local approximation \cite{ref2}. Operation under atmospheric pressure is planned for future experiments to extend the range of applications to the biomedical and environmental field \cite{applicationatm1, application5}. Parallel to the experimental studies the theoretical description of the jet has progressed \cite{ref4}. The work presented by Klute et al. is based on two submodels: First, an electromagnetic model for the fields $\vec{E}$, $\vec{B}$ and the microwave power $P_{\mathrm{abs}}$, absorbed by the plasma. And second, a global model for an argon plasma. The resulting description is in good agreement with some experimental findings but it can not claim quantitative validity as the plasma model includes strong simplifications and is only available for argon. The purpose of this work is to correct these deficiencies and present a quantitatively accurate model for nitrogen at $p =1000\,\mathrm{Pa}$. This is achieved by combining the electromagnetic part of \cite{ref4} with a global chemistry model for nitrogen, based on Thorsteinsson and Gudmundsson \cite{ref5}, customized for the geometry of the MMWICP. The paper is structured as follows: Section II presents the electromagnetic field model as well as the global chemistry model, including the necessary modifications for the MMWICP. Both submodels are coupled self-consistently. Section III briefly outlines the diagnostics used for the experimental characterisation. The results of the theoretical model are compared with experimental data in section IV. In addition, the balance of capacitive and inductive coupling in the absorbed power is investigated for different discharge regimes. The paper concludes with a summary and conclusion in section V.

\newpage
\section{Theoretical model}
\subsection{Electromagnetic model}
\begin{figure}[h]
\centering
\includegraphics[width=0.7\textwidth]{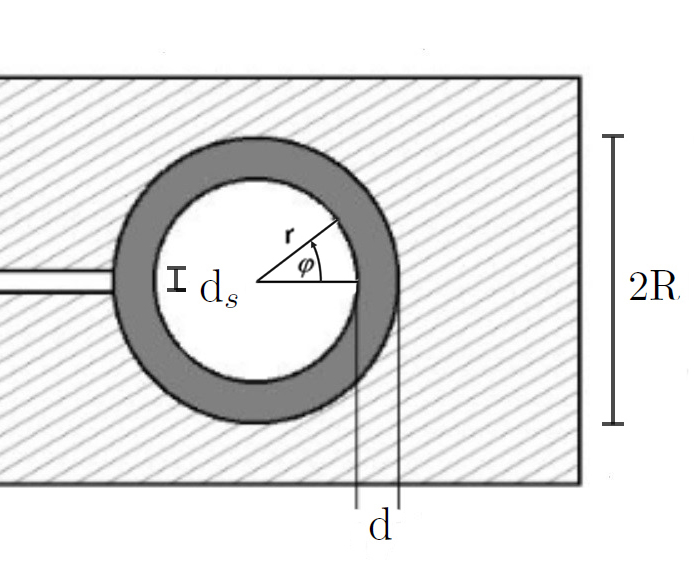}
\begin{flushleft}
\caption{Cross section and inner dimensions of the resonator, where $\mathrm{R}$ represents the radius of the borehole, 
	 $\mathrm{d}$ the wall thickness of the dielectric tube and $\mathrm{d}_\mathrm{s}$ the width of the capacitor gap. The coordinate system used for calculation is shown inside the cavity.}
\label{fig:bench}
\end{flushleft}
\end{figure}

The model of Klute et al. is used for the electromagnetic description of the MMWICP \cite{ref4}. It uses cylindrical coordinates $(r,\phi,z)$ with natural orientation as shown in Fig. \ref{fig:bench}. With respect to the variable $r$ and the cavity's radius $R$, there are three different zones defined as follows: The inner zone from $r=0$ to $r=R-d-\delta$ contains plasma of a constant electron density $n_\mathrm{e}$, the zone from $r=R-d-\delta$ to $r=R-d$ is the electron-depleted sheath (with constant length $\delta$ in this model), and the zone from $r=R-d$ to $r=R$ is the dielectric tube with thickness $d$. The magnetic field $\vec{B}$ and the electric field $\vec{E}$ are assumed to be invariant in the $z$-direction, but depend on $r$ and $\phi$. This is due to the capacitor gap at $\phi=0$, which breaks the azimuthal symmetry. Using a time harmonic approach, the fields for the plasma zone, which is of interest to this work, are written as follows:
\begin{align}
\vec{B}(r,\phi,t)&=\mathrm{Re}\left(\underline{B}_z(r,\phi)\exp(i\omega t)\vec{e}_{z}\right), \\[0.5ex] 
\vec{E}(r,\phi,t)&=\mathrm{Re}\left(\underline{E}_r(r,\phi)\exp(i\omega t)\vec{e}_{r}+\underline{E}_{\phi}(r,\phi)\exp(i\omega t)\vec{e}_{\phi}\right),
\end{align}
with the additional charge density $\rho$ and the current density $\vec{j}$:
\begin{align}
  \rho(r,\phi,t)  &= \mathrm{Re}\Bigl(\underline{\rho}(r,\phi)\exp(i\omega t)\Bigr), \\[0.5ex] 
	\vec{j}(r,\phi,t)&=\mathrm{Re}\left(\underline{j}_r(r,\phi,t)\exp(i\omega t)\vec{e}_{r}+\underline{j}_{\phi}(r,\phi)\,\exp(i\omega t)\vec{e}_{\phi}\right).
\end{align}
The full set of Maxwell equations,
\begin{align}
	  &\frac{1}{\mu_0}\frac{1}{r}\frac{\partial \underline{B}_z}{\partial \phi} = \underline{j}_r+\varepsilon_0 i \omega \underline{E}_r\label{eq:221}, \\[0.1ex]
	 	&	-\frac{1}{\mu_0}\frac{\partial \underline{B}_z}{\partial r} = \underline{j}_{\phi}+\varepsilon_0 i\omega \underline{E}_{\phi}^{(\mathrm{p})},\\[0.1ex]
    & \frac{1}{r}\frac{\partial(r \underline{E}_{\phi}) }{\partial r}-\frac{1}{r}\frac{\partial \underline{E}_r}{\partial \phi} 
	= -i\omega \underline{B}_z,
\end{align}
is considered and coupled to the cold plasma model. It comprises of the equation of charge conservation
\begin{align}
	i\omega\underline{\rho}+ \frac{1}{r}\frac{\partial(r \underline{j}_r) }{\partial r}
	    +\frac{1}{r}\frac{\partial \underline{j}_{\phi}}{\partial \phi}=0,
\end{align}
and the equation of motion, where $\omega_\mathrm{pe}$ is the plasma frequency and $\nu$ the 
collision rate,
\begin{align} 
	   &i\omega \underline{j}_r=  \varepsilon_0\omega_\mathrm{pe}^2 \underline{E}_r-\nu \underline{j}_r,\\
	   &i\omega \underline{j}_{\phi}=\varepsilon_0\omega_\mathrm{pe}^2 \underline{E}_{\phi}-\nu \underline{j}_{\phi}\label{eq:222}.
\end{align}
Second order differential equations arise for $\vec{E}$ and $\vec{B}$. They are analytically solved in the frequency domain. A Fourier-series approach is used to account for the symmetry breaking due to capacitor gap. The solutions consist of an infinite number of modes ordered by the azimuthal wavenumber m:
\begin{align}\label{eq:1}
	\underline{B}_z(r,\phi)&=u\sum_{m=0}^{\infty}C_m
	J_m\left(\sqrt{\varepsilon_\mathrm{p}}\;\frac{\omega}{c}r\right)\cos(m\phi),\\
\vec{E}(r,\phi)&=\mathrm{Re}\left(\frac{c^2}{i\omega\varepsilon_\mathrm{p}}\frac{1}{r}\frac{\partial\underline{B}_z}{\partial  \phi}\vec{e}_{r} -\frac{c^2}{i \omega\varepsilon_\mathrm{p}}\frac{\partial \underline{B}_z}{\partial r}\vec{e}_{\phi}\right),   \label{eq:12}     	
\end{align}
where the $J_{m}$ denotes Bessel functions of the first kind and order $m$, $\varepsilon_{\mathrm{p}}= 1-\frac{\omega_{\mathrm{pe}}^2}{\omega^2 -i \omega\nu}$ the relative plasma permittivity, $c$ the speed of light and $\omega$ and $u$ the driving frequency and amplitude voltage of the microwave signal at the gap capacitor. The constants $C_{m}$ can be found in \cite{ref4}. The expressions \ref{eq:1} and \ref{eq:12} contain the spatial resolution of the model. Evaluating the discontinuity condition for the magnetic field at the cavity boundary, the admittance of the plasma can be derived \cite{ref4}:
\begin{align}
  Y_{\mathrm{p}}(\omega,n_{\mathrm{e}})=\sum_{m=0}^{\infty} Y_{m}(\omega,n_{\mathrm{e}}).
\end{align}
The mode $m$=$0$ is related with inductive coupling and the modes $m\geq1$ with capacitive coupling \cite{ref4}. This allows to cast the system of plasma, resonator and matching network in the form of a lumped element equivalent circuit \cite{ref4}. With the resulting admittance for the combined system, $Y_{\mathrm{s}}(\omega,n_{\mathrm{e}})$, the power absorbed by the plasma reads
\begin{align}
	  P_{\mathrm{abs}}(\omega,n_{\mathrm{e}}) =
	  \frac{1}{2}  \mathrm{Re} \left(Y_{\mathrm{s}}(\omega,n_{\mathrm{e}}) \right) u_{\mathrm{s}}^2, 
\end{align}
where $u_{\mathrm{s}}$ represents the voltage of the microwave generator. As the original work was done for argon and $p=100\,\mathrm{Pa}$, the higher collision rate must be corrected. For a gas particle density of $n_{\textrm{g}}=4.65 \!\times\!10^{22} \, \mathrm{m}^{-3}$, a cross section of $2.5 \!\times\!10^{19} \, \mathrm{m}^{2}$ \cite{ref7} and a thermal electron velocity of $6.73 \!\times\!10^{5} \, \mathrm{m/s}$ (for a measured gas temperature of $T_{\mathrm{g}}=1600\,\mathrm{K}$), the collision rate is given by $\nu=n_{\textrm{g}}\sigma v_{\textrm{th}}=7.8\times10^{9}\,\textrm{Hz}$.
\newpage

\subsection{Plasma model}
The plasma is represented via a volume-averaged \textit{global} model developed by Thorsteinsson and Gudmundsson \cite{ref5}. The model is briefly described here and further details are provided by the original study. Our model implementation is subjected to a code-to-code verification with the simulation results, as well as the validation against the measurements presented in \cite{ref5} and a good agreement is obtained. The chemical kinetics properly addresses the considered pressure regime (e.g. compared to high pressure models of Sakiyama et al. \cite{Sakiyama_2012}) and additional modifications are omitted.
\iffalse
% Here is the previous version
The plasma is represented via a volume-averaged \textit{global model} developed
by Thorsteinsson and Gudmundsson \cite{ref5}. The model is briefly described here and further details
are provided by the original study. 
\fi
A total of 15 
nitrogen species are included: 
The seven lowest vibrationally excited states of the nitrogen molecule in the ground state $\textrm{N}_{2}(X^1\Sigma_{g}^{+},v=0-6)$, the metastable molecule $\textrm{N}_{2}(\textrm{A}^3 \Sigma_{\textrm{u}}^+)$, the ground state atom $\textrm{N}(^4\textrm{S})$, the metastable atoms $\textrm{N}(^2\textrm{D})$ and $\textrm{N}(^2\textrm{P})$ as well as the ions $\textrm{N}^+$, $\textrm{N}_{2}^+$, $\textrm{N}_{3}^+$ and $\textrm{N}_{4}^+$. Every species $X$ follows the volume-averaged balance equation  
\begin{align}
\frac{\textrm{d}n^{(X)}}{\textrm{d}t}=\sum_{i}R_{\textrm{gen},i}^{(X)}-\sum_{i}R_{\textrm{loss},i}^{(X)},
\end{align}
where $n^{(X)}$ denotes the particle density, $R_{\textrm{gen},i}^{(X)}$ represents a net 
generation process and $R_{\textrm{loss},i}^{(X)}$ a net loss process. The volume-averaged 
energy balance is written as
\begin{align}\label{eq:2}
\frac{\textrm{d}}{\textrm{d}t}\left(\frac{3}{2} e n_{\mathrm{e}} T_{\mathrm{e}}\right)=\frac{1}{V}\left(P_{\textrm{abs}}
-P_{\textrm{che}}
%-e V n_{\mathrm{e}} \sum_{\alpha}n_{\alpha}\epsilon_{c \alpha}k_{iz \alpha}
-e u_{\textrm{B}}n_{\textrm{i}}A_{\textrm{eff}}(\epsilon_{\textrm{i}}+\epsilon_{\textrm{e}})\right),
\end{align} 
where $P_{\mathrm{abs}}$ is the power gained from the electromagnetic fields by the electrons and $P_{\textrm{che}}$ is the net electron energy loss in the homogeneous chemical reactions. The third term on the right side accounts for the energies $\epsilon_{i}$ and $\epsilon_{e}$, carried off by the ions and electrons lost to the mantle surface with the effective area $A_{\textrm{eff}}=2 \pi h_R(R-d) L = h_{R}\,125\, \mathrm{mm}^2$, where $e$ is the elementary charge, $u_{\textrm{B}}$ the Bohm velocity and $n_{\textrm{i}}$ the ion density. The edge to center scalig factor $h_{R}$ is given by
\begin{align}
h_{R} \approx 0.80\left[4+\frac{R-d}{\lambda_{\mathrm{i}}}+\left(\frac{0.80 (R-d) u_{\mathrm{B}}}{\chi_{01} J_{1}\left(\chi_{01}\right) D_{\mathrm{a}}}\right)^{2}\right]^{-1 / 2},
\end{align}
where $\lambda_{\mathrm{i}}$ is the mean free path of ions, $\chi_{01}\approx2.405$ is the first zero of the zero order Bessel function $J_{0}$ and $D_{\mathrm{a}}$ is the ambipolar diffusion coefficient \cite{globalmodel2}. Only radial losses are considered, since there are no axial boundaries and the plasma column extends beyond the resonator. The sum of the two loss terms in (\ref{eq:2}) is defined as $P_{\textrm{loss}}(n_{\mathrm{e}})$, representing the total power loss to the plasma. Finally, the volume of the discharge is denoted with $V=\pi (R-d)^2 L$. The model is evaluated for an equilibrium state. A pressure of $p =1000\,\mathrm{Pa}$ is assumed here; the gas temperature is linearly interpolated between $T_{\mathrm{g}}=650\,\mathrm{K}$ and $T_{\mathrm{g}}=1600\,\mathrm{K}$, depending on the power. The electron density follows from the particle balance as well as quasi neutrality and the electron temperature from the energy balance. 
%The validity of global models for the considered pressure range is verified in \cite{ref10,ref11} and \cite{ref12}.
The authors refer to earlier studies \cite{ref10,ref11,ref12} for a validation and a verification of the global model of microwave-induced discharges in the considered pressure regime.

%The model has been benchmarked as shown in figures \ref{fig:bench}. 
%"The global model of microwave induced discharges in the considered pressure regime are verified \cite{Kem15} and validated \cite{Kem17,Kem19}." see the commented out links for the references.
% Kem15 : https://doi.org/10.1088/0022-3727/48/43/435203
% Kem17 : https://doi.org/10.1088/1361-6463/aa7093

\newpage
\color{black}
%%%%%%%%%%%%%%%%%%%%%%%%%%%%%%%%%%%%%%%%%%%%%%%%%%%%%%%%%%%%%%%%%%%%%%%

\section{Diagnostics}
The diagnostics used for the experimental characterisation of the MMWICP are briefly outlined below. The used methods can be categorized as either optical or microwave based.

\subsection{Optical measurements}
\begin{figure}[h]
\centering
\includegraphics[width=0.80\textwidth]{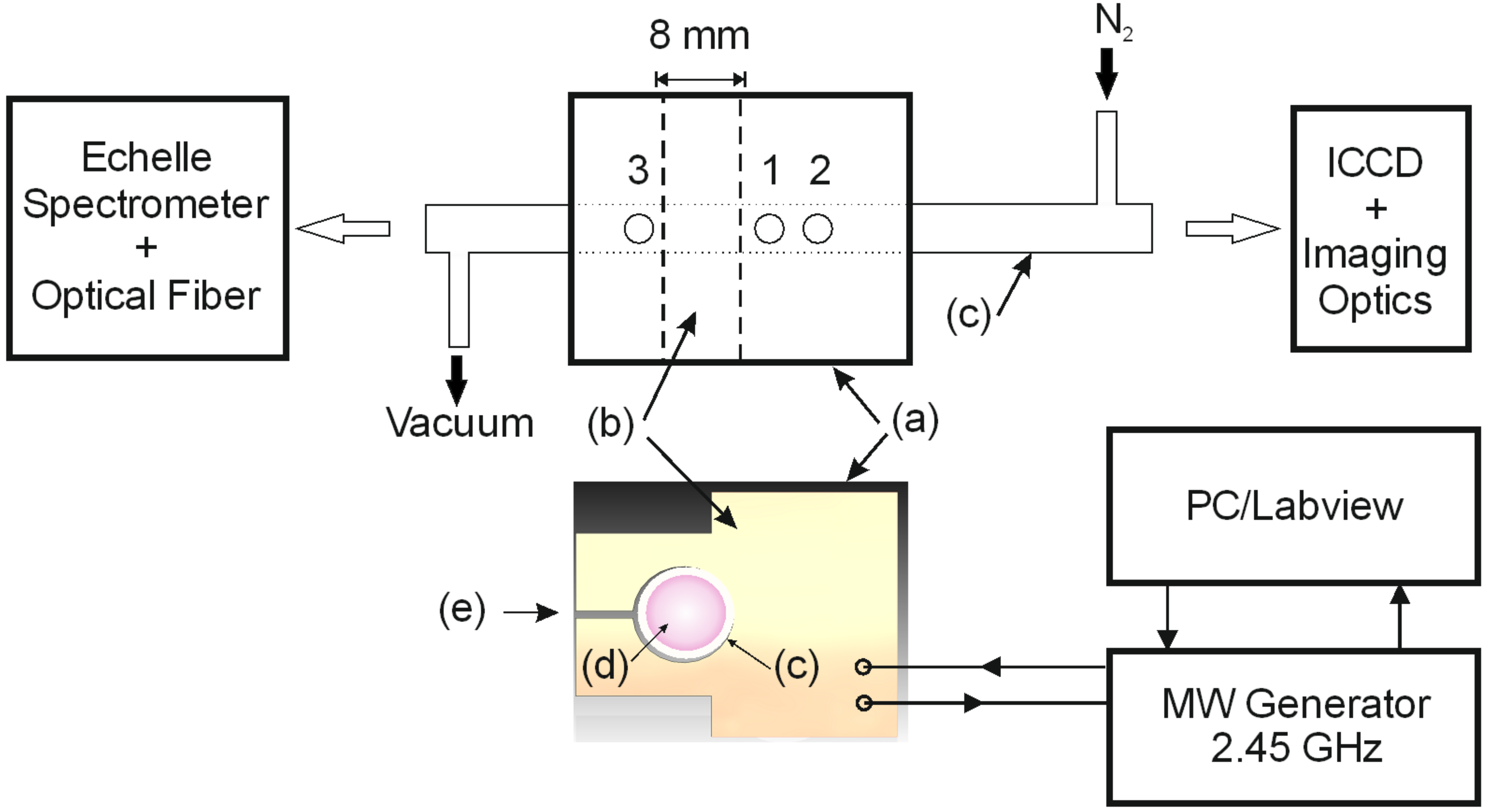}
\begin{flushleft}
\caption{Experimental set-up used for the optical and spectroscopical characterization of the MMWICP. Top: Optical set-up. Bottom: Cross section of the resonator perpendicular to the optical axis. (a) Al-shielding, (b) resonator, (c) quartz tube, (d) plasma, (e) capacitor gap.}
\label{fig:Aufbau}
\end{flushleft}
\end{figure}

Optical and spectroscopical measurements are performed using the experimental set-up shown in Fig. \ref{fig:Aufbau}. High-resolution optical emission spectroscopy (HROES) is used to determine global plasma parameters, since the Langmuir probe is not applicable for the MMWICP. In particular the UV system of Nitrogen is investigated. A high dispersion Echelle spectrometer (resolution of $\Delta\lambda=0.015-0.06\,\textrm{nm}$ for $\lambda=200-800\,\textrm{nm}$) provides the rotational distributions of the second positive (N$_{\mathrm{2}}$(C-B)) and the first negative (N$_\mathrm{2}^+$(B-X)) nitrogen system. A theoretical model is fitted to the measured spectra by varying the assumed ion and gas temperature \cite{ref2}. This allows the determination of the neutral gas temperature $T_\textrm{g}$. To calculate the electron density $n_\mathrm{e}$, the plasma kinetics will be described in a local approximation. (This is well fulfilled for our experimental conditions at $p =1000\,\mathrm{Pa}$.) A Boltzmann solver then provides the relation between the reduced electrical field and the electron energy distribution function \cite{ref8}. By evaluating a simple collision radiation model, the (global) electron density $n_\mathrm{e}$ can be calculated. All calculations are performed for an equilibrium state and are based on known cross sections for the according processes \cite{ref7}. A detailed description for this method can be found in \cite{ref6} and \cite{ref20}.
\vspace*{3mm}\newline 
An imaging system is used to investigate the spatial distribution of the emission. The setup consists of a telecentric macro lens, an ICCD camera with a resolution of 28\,µm and two alternating Fabry-Perrot filters in front of the lens (to select the spectral lines of interest). By adjusting the filters to either $380\,\textrm{nm}$ or $391\,\textrm{nm}$, the emission of the second positive (N$_\mathrm{2}$(C-B)) and the first negative (N$_\mathrm{2}^+$(B-X)) nitrogen system are captured spatially resolved. By comparison with the simultaneously measured results of the Echelle spectrometer, the intensities can be determined absolutely. The collision radiation model then provides the strength and spatial distribution of the electric field $\vec{E}$. In addition to the optical measurements, the signal generator is used to determine the incident and reflected power.

\newpage
\subsection{Microwave measurements}
\begin{figure}[h]
\centering
\includegraphics[width=0.80\textwidth]{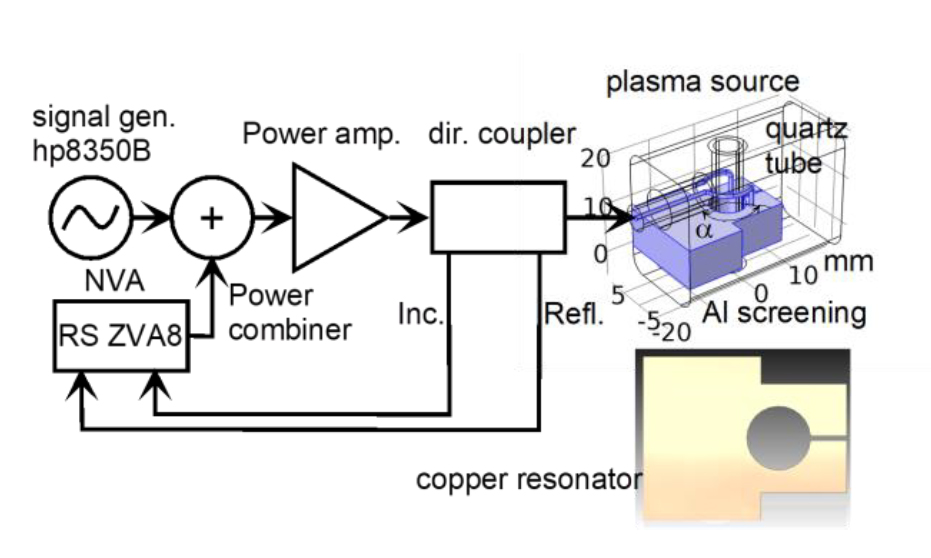}
\begin{flushleft}
\caption{Experimental set-up for the \grq Hot-S-parameter' spectroscopy. This measurement principle enables the simultaneous excitation of the plasma and measurement of the complex $S_{11}$ parameter.}
\label{fig:HotS}
\end{flushleft}
\end{figure}

The microwave diagnostics of the MMWICP are based on the so called \grq Hot-S-Parameter' spectroscopy, described in detail in \cite{ref3}. For this method two microwave signals are combined as shown in Fig. \ref{fig:HotS}: First, a signal from a microwave generator (hp8350B) and second, from a network analyzer (Rohde \& Schwarz, ZVA8). The strong microwave signal with fixed power and fixed frequency of $2.45\,\textrm{GHz}$ excites the plasma and defines the parameters of the discharge. The second, much weaker microwave signal ($-30\,\textrm{dB}$) with varying frequency is overlapped in order to investigate the plasma. The sum signal is amplified and sent to the plasma source. It is assumed that no significant harmonics and intermodulation products occur and the plasma remains unaffected by the probe signal. This applies in microwave frequencies since the recombination time of electrons and ions is in the range of microseconds (much longer than a microwave period). For a fixed set of plasma parameters ($n_\mathrm{e}$, $T_\mathrm{g}$,..), the frequency of the probe signal will be varied. The network analyzer then provides the complex scattering parameter $S_{11}$ as a function of the frequency. The results can be represented as resonance curves in the frequency domain. In addition the plasma impedance and coupling efficiency can be derived \cite{ref3}.

\newpage

\section{Results}
\subsection{Global plasma parameters}

\begin{figure}[ht]
	\centering
		\includegraphics[width=0.700\textwidth]{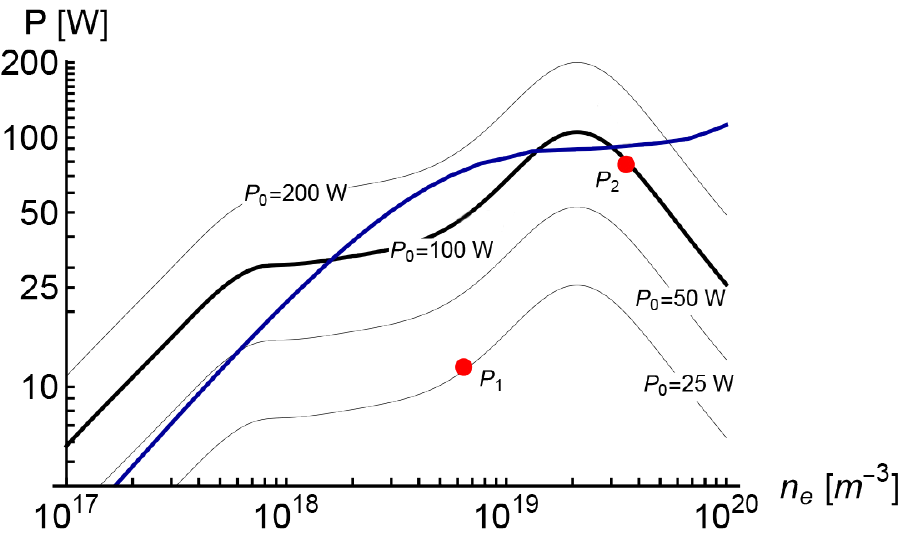}
		\begin{flushleft}
	\caption{Absorbed power $P_{\mathrm{abs}}$ (black curves) and loss power $P_{\mathrm{loss}}$ (blue curve) from the theoretical model as a functions of 
    $n_{\mathrm{e}}$. $P_{\mathrm{abs}}$ is shown for four different incident powers $P_0$. The intersections of both curves represent stationary points of the model. The experimentally determined points have been marked with $\textrm{P}_1$ and $\textrm{P}_2$.}
    \label{fig:2}
    \end{flushleft}
\end{figure}

\begin{figure}[ht]
	\centering
		\includegraphics[width=0.700\textwidth]{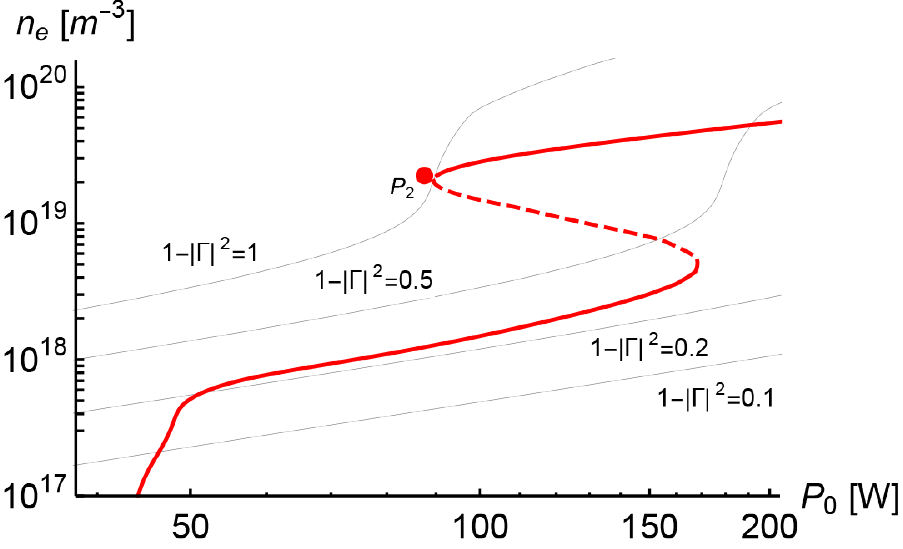}
		\begin{flushleft}
	\caption{Stable (solid) and unstable (dashed) stationary points in the plane $P_0$ (incident power) and $n_{\mathrm{e}}$ (electron density). $1-\left|\Gamma\right|^2$ represents the fraction of $P_0$ that is absorbed by the plasma. The experimentally determined point $P_{2}$ has been marked.}
    \label{fig:Hysterese}
    \end{flushleft}
\end{figure}

This section compares results from the theoretical model with experimental data. First, volume averaged (global) parameters are considered. Fig. \ref{fig:2} shows the absorbed power $P_{\mathrm{abs}}(\omega,n_{\mathrm{e}})$ as a function of $n_\mathrm{e}$ for various values of the incident microwave power $P_0$. The curves of $P_{\mathrm{abs}}(\omega,n_{\mathrm{e}})$ show two peaks. The first peak appears at low electron densities and results from the resonant capacitive modes $m$$\geq$$1$; it represents the capacitive regime. The second peak appears at high values of $n_\mathrm{e}$ and results from the mode $m$=$0$; it represents the inductive regime. The second (blue) curve in Fig. \ref{fig:2} is the loss power $P_\textrm{loss}(n_\mathrm{e})$, calculated from the global plasma model. Intersections of the curves represent the stationary points of the model. Fig.~\ref{fig:Hysterese} shows the system characteristics in the  $P_0$-$n_\mathrm{e}$-plane (stable stationary points are represented by the solid lines; unstable stationary points by the dashed lines). The experimentally determined plasma parameters are listed in Table 1 and have been marked in Fig. \ref{fig:2} as $P_{1}$ and $P_{2}$ \cite{ref2}. Simulation and experiment both show two stable operating regimes, separated by a significant change of the plasma parameters and coupling efficiency. The measured and calculated stationary points in the $n_\mathrm{e}$-$P_{\mathrm{abs}}$-plane show a remarkable good agreement for the high $n_\mathrm{e}$ case, whereas model and experiment significantly differ for the low $n_\mathrm{e}$ case. This can possibly be explained by the spatial inhomogeneity of the discharge at low absorbed powers, as it follows from the results in section D. Due to its nature as volume averaged model, the global model assumes a homogeneous electron density and electron temperature in the entire volume. These conditions are well fulfilled for the homogeneous H-mode, but not if the plasma is operated at low absorbed power. As a consequence the loss power is distorted to unrealistically high values in the latter case.

\begin{table}[ht]
\begin{tabular}{l*{2}{c}r}
\hline
$P_{0}$(W) &
 $P_{\mathrm{abs}}$(W)            & $T_{\mathrm{g}}$(K)  & $n_{\mathrm{e}}$ ($\mathrm{m}^{-3}$)  \\
\hline
 \,\,\,$20$ & $12$ & $650\pm20$ &                    $(6.4\pm2.7)\times 10^{18}$  \\
 \,\,\,$88$ & $78$ & $1600\pm100$ &                    $(3.5\pm1.7)\times 10^{19}$   \\
\hline
\label{tab:caption}
\end{tabular}
\caption{\label{tab:table1} Results of the experimental characterization.}
\end{table}
\subsection{Microwave measurements}
\begin{figure}[h]
	\centering
		\includegraphics[width=1.00\textwidth]{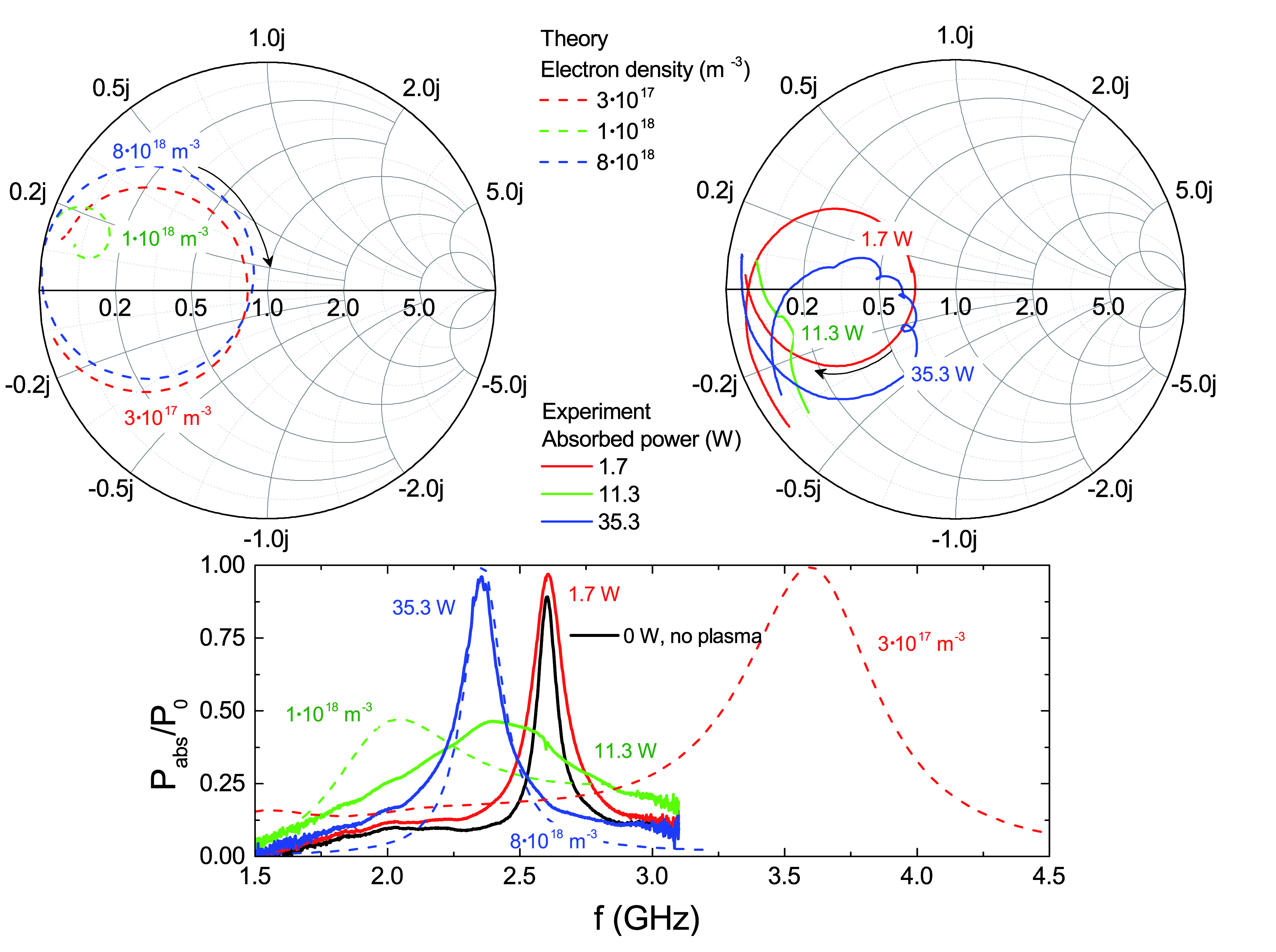}
		\begin{flushleft}
	\caption{Top: Smith chart representation of the complex impedance as a function of the frequency $\omega$ for different electron densities (theory, left) and different excitation powers (experiment, right). Arrows indicate the variation of frequency from low to high. The considered frequency range is identical to the lower figure. The experimental curves have been rotated by a phase shift to show only the contribution of the resonator. Bottom: Ratio of the absorbed power to the incident power $P_{\textrm{abs}}/P_{0}=1-|S_{11}|^2$ as a function of the frequency $\omega$ for different electron densities (theory, dashed lines) and different excitation powers (experiment, solid lines).
 }
\label{Smithchart}
\end{flushleft}
\end{figure}

In order to compare the microwave measurements with the model, the theoretical $S_{11}$-parameter is calculated using the transformation $S_{11}$=$(Y_\mathrm{s}(\omega,n_{\mathrm{e}})^{-1}-50\,\Omega)/(Y_\mathrm{s}(\omega,n_{\mathrm{e}})^{-1}+50\,\Omega)$. By varying the excitation frequency $\omega$ for a fixed electron density, the evolution of the $S_{11}$-parameter in the frequency domain can be studied. The resulting parametric plots and their experimental correspondents are represented as resonance circles in a Smith chart (upper part of Fig. \ref{Smithchart}). A good correlation between experiment and theory can be observed in the frame of these results: With increasing electron density and absorbed power respectively, the resonance circles in Fig. \ref{Smithchart} are contracting and approach the point $z$=$0$ (left in the Smith chart) i.e., the plasma conductivity is increasing and the impedance is decreasing. The impedance reaches its minimum (green line) in the capacitive regime. With further increasing electron density and the absorbed power respectively, the plasma conductivity increases, but also the impedance as the conductive zone is decreasing due to the skin effect. The resonance circles are expanding again and approach the point $z$=$1$ (center of the Smith chart) i.e., the coupling efficiency is improving as the point $z$=$1$ represents ideal impedance matching. This behavior indicates the onset of the inductive coupling \cite{ref3,ref4}. The lower part of Fig. \ref{Smithchart} shows the ratio of the absorbed power to the incident power $P_{\textrm{abs}}/P_{0}=1-|S_{11}|^2$ as a function of the excitation frequency $\omega$. The agreement between theory and experiment depends again on the absorbed power and electron density. In the case of high absorbed power (blue lines), the experimental data is accurately reproduced, showing a coupling efficiency of $\geq60\,\%$. The capacitive case at a medium value of electron density and absorbed power shows as expected, a low coupling efficiency of $\leq50\,\%$ and discrepancies between model and experiment. At very low electron densities (red line) the resonance curves approach the case without plasma. This causes significant deviations between theory and experiment, presumably due to the idealized description of the electrical network in the model.

%For the high $n_{\mathrm{e}}$/$P_{0}$ case the frequency dependent behaviour is accurately reproduced, showing a coupling efficiency $\geq60\%$.
%The radius of the resonance circles is correlated with the realpart of the impedance.
% and allows an ever better energy transfer to plasma due to matching with the generator impedance
%In terms of lumped elements the plasma is represented as a series circuit of an LR branch and a several LCR branches \cite{ref4}. 
%$S_{11}$ is evaluated for a fixed $n_\mathrm{e}$ and as a function of $\omega$. 
%The experimental power range only extends up to 35W.
\subsection{Mode analysis}
\begin{figure}[h]
	\centering
		\includegraphics[width=0.80\textwidth]{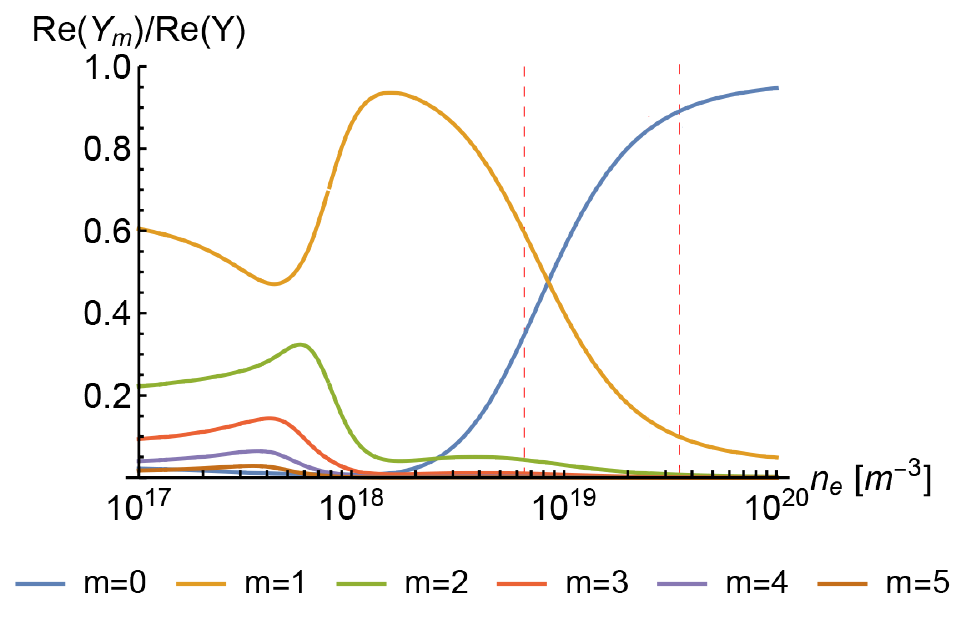}
		\begin{flushleft}
	\caption{Ratio of the individual modes $Y_\mathrm{0}(\omega,n_{\mathrm{e}})..Y_\mathrm{5}(\omega,n_{\mathrm{e}})$ to $Y_\mathrm{p}(\omega,n_{\mathrm{e}})$ (real parts) as a function of $n_{\mathrm{e}}$. The experimentally determined electron densities $n_\mathrm{e}=6.4\!\times\!10^{18} \, \mathrm{m}^{-3}$ (first dashed line) and $n_\mathrm{e}=3.5\!\times\!10^{19} \, \mathrm{m}^{-3}$ (second dashed line) have been marked.}
\label{modenanalyse}
\end{flushleft}
\end{figure}

The results are complemented by a mode analysis: Based on the theoretical model the share of the individual modes $Y_\mathrm{1}(\omega,n_{\mathrm{e}}),Y_\mathrm{2}(\omega,n_{\mathrm{e}}),...Y_\mathrm{m}(\omega,n_{\mathrm{e}})$ in $Y_\mathrm{p}(\omega,n_{\mathrm{e}})$ will be examined for different stationary situations (only the real part of Y is considered). This provides insights into the physical character of the discharge. Fig. \ref{modenanalyse} shows $\mathrm{Re}(Y_\mathrm{m}(\omega,n_{\mathrm{e}})) / \mathrm{Re}(Y_\mathrm{s}(\omega,n_{\mathrm{e}}))$ for $m$=$0,1,...5$ and as a function of $n_\mathrm{e}$. The experimentally determined stationary points have been marked. For $n_\mathrm{e}=6.4\!\times\!10^{18} \, \mathrm{m}^{-3}$ the modes $m$=$0$ and $m$=$1$ are dominating. The corresponding discharge should be therefore called E/H-hybridmode, as it includes a strong capacitive and inductive component. This was first proposed in \cite{ref2}. For $n_\mathrm{e}=3.5\!\times\!10^{19} \, \mathrm{m}^{-3}$ the inductive $m$=$0$ mode is dominating and confirms the interpretation of this discharge as H-mode. The cases discussed in the microwave section can be characterized as follows: For $n_\mathrm{e}=1\!\times\!10^{18} \, \mathrm{m}^{-3}$ the mode $m$=$1$ is dominating, confirming the pure capacitive character of the discharge. For $n_\mathrm{e}=8\!\times\!10^{18} \, \mathrm{m}^{-3}$ the capacitive $m$=$1$ and inductive $m$=$0$ modes make roughly equal contributions, confirming the interpretation of this case as onset of the inductive coupling.

\subsection{Spatially resolved results}
\begin{figure}[h]
\centering
\includegraphics[width=1.00\textwidth]{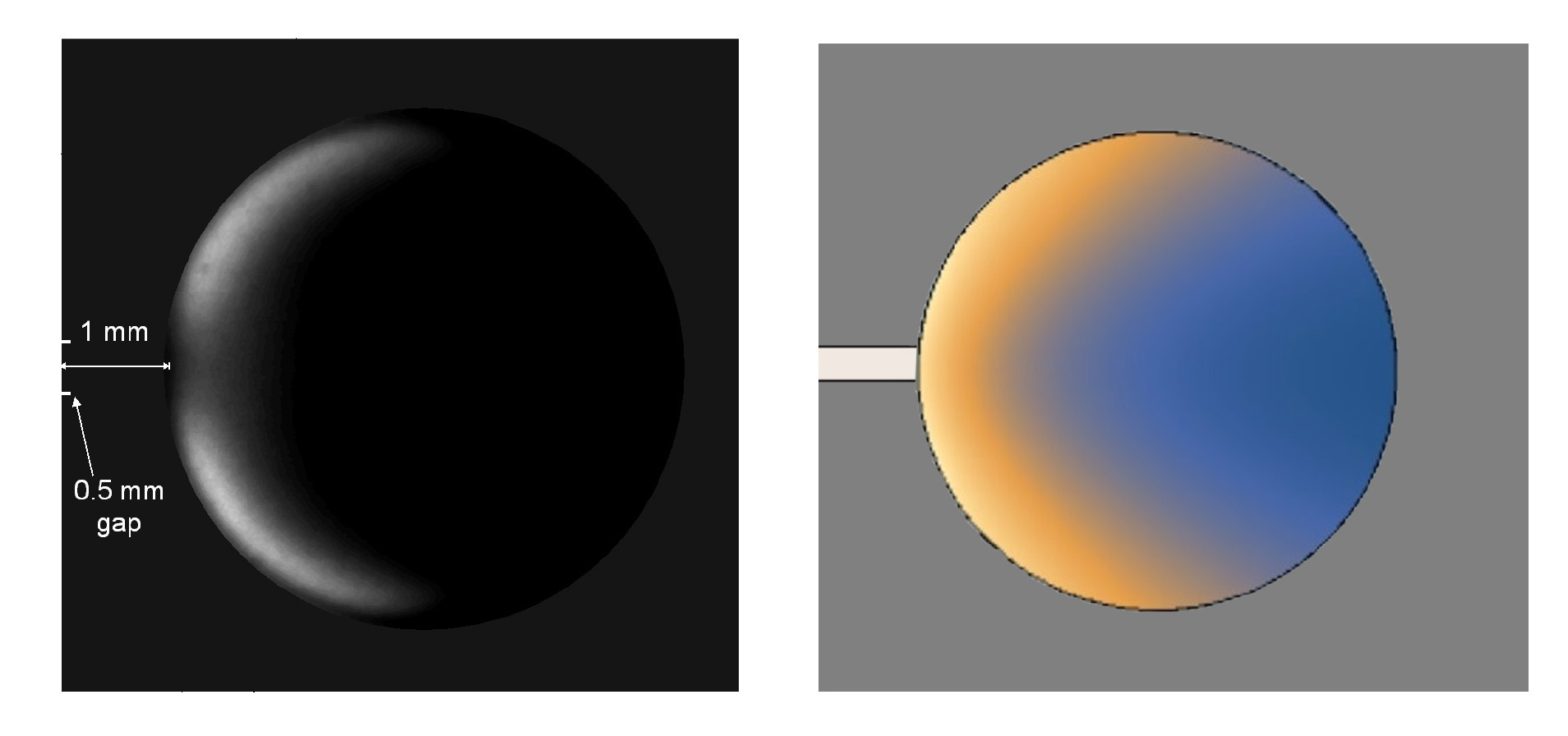}
\begin{flushleft}
\caption{Radial image of the plasma zone for $P_{\mathrm{abs}}=12\, \mathrm{W}$ (left) and distribution of $|\vec{E}|^2$ as calculated from the theoretical model for $n_\mathrm{e}=6.5\!\times\!10^{18} \, \mathrm{m}^{-3}$ (right).}
\label{fig:EModeImage}
\end{flushleft}
\end{figure}

\begin{figure}[h]
	\centering
		\includegraphics[width=1.00\textwidth]{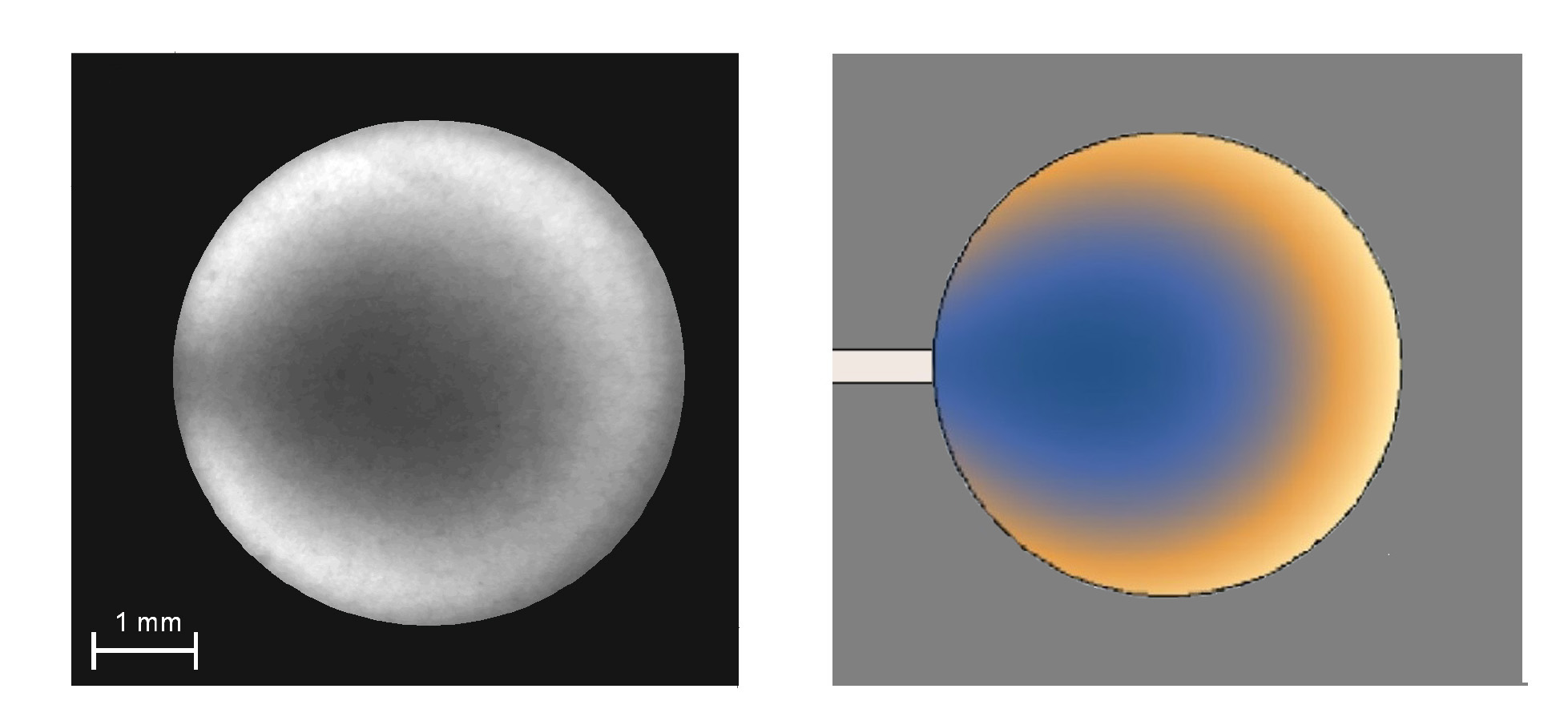}
	\caption{Radial image of the plasma zone for $P_{\mathrm{abs}}=78\, \mathrm{W}$ (left) and distribution of $|\vec{E}|^2$ as calculated from the theoretical model for $n_{\mathrm{e}}=3.5\!\times\!10^{19} \, \mathrm{m}^{-3}$ (right).}
	\label{fig:HModeImage}
\end{figure}

\begin{figure}[h]
\centering
\includegraphics[width=1.00\textwidth]{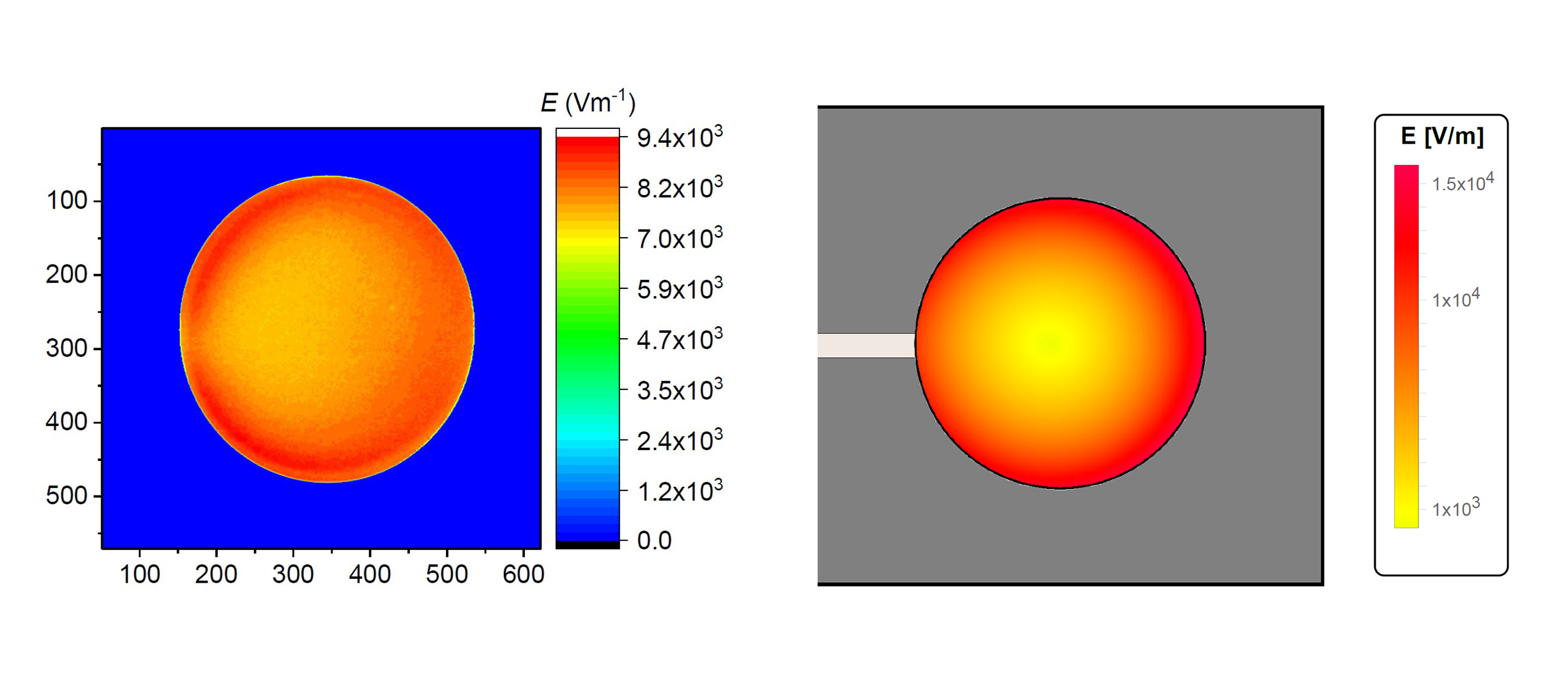}
\begin{flushleft}
\caption{Measured electric field strength in the $r$-$\phi$-plane for $P_{\mathrm{abs}}=78\, \mathrm{W}$ (left) and simulated electric field strength $|\vec{E}|$ as calculated from the model for $n_\mathrm{e}=3.5\!\times\!10^{19} \, \mathrm{m}^{-3}$ (right).}
\label{fig:EFieldImage}
\end{flushleft}
\end{figure}

 The left part of Figs. \ref{fig:EModeImage} and \ref{fig:HModeImage} present the light emissions in the $r$-$\phi$-plane as seen from a perpendicular view and for two different stationary situations. For a qualitative comparison with the theoretical model, 2-dimensional plots of $|\vec{E}|^2$ are used as shown in Figs. \ref{fig:EModeImage} and \ref{fig:HModeImage} on the right. This is reasonable, since in the local regime the light emission intensity is approximately proportional to the local absorbed power $\langle\vec{j}\cdot\vec{E}\rangle\propto|\vec{E}|^2$. Figs. \ref{fig:EModeImage} and \ref{fig:HModeImage} both show characteristic emission patterns and a good correlation between simulation and experiment. At low absorbed power a narrow emission zone is formed close to the gap capacitor (Fig. \ref{fig:EModeImage}). At high absorbed power the emission zone resembles a ring shape and appears comparatively homogeneous (Fig. \ref{fig:EModeImage}). Fig. \ref{fig:EFieldImage} shows the measured and calculated electric field strength $|\vec{E}|$ in the $r$-$\phi$-plane. Simulation and measurements show matching spatial distributions, only the radial gradient of the is more pronounced in the simulation. Quantitative agreement can also be determined within typical limits. In conclusion the morphology of the discharge strongly depends on the absorbed power and electron density respectively. This supports the discussion in section A, which explains deviations between model and theory at low powers as a consequence of the spatial inhomogeneity.

\newpage
\section{Summary and conclusion}
In this work we employed an existing electromagnetic model for the MMWICP and coupled it self consistently to a global model for a nitrogen plasma. It results a quantitative description for a non-thermal nitrogen plasma at a pressure of $p = 1000\,\mathrm{Pa}$ and a gas temperature of $T_\mathrm{g}=650\mbox{-}1600\,\mathrm{K}$. This goes beyond previous modelling and enables a comparison with experimental data, including volume averaged plasma parameters, the spatial distribution of light emissions and the electrical field strength and microwave measurements. Simulation and experiment both show two stable stationary points, separated by a significant change of the plasma parameters and coupling efficiency. For the measured and calculated stationary point at high absorbed power $P_{\mathrm{abs}}$ and high electron density $n_\mathrm{e}$ there is a remarkable correspondence in the $n_\mathrm{e}$-$P_{\mathrm{abs}}$-plane, whereas model and experiment significantly differ for low values of $P_{\mathrm{abs}}$ and $n_\mathrm{e}$ respectively. This is possibly a consequence of the spatial inhomogeneity of the discharge in the latter case as the global model cannot capture such a situation well. The spatial distribution of the local absorbed power $\langle\vec{j}\cdot\vec{E}\rangle$ is found to be in good correlation with the measured light emission for both discharge regimes. As expected, the discharge shows a strong spatial inhomogeneity in the low power regime. The measured and calculated distribution of $|\vec{E}|$ also match within typical limits. Another finding is based on a mode analysis of the plasma impedance: The experimental stationary point at $n_\mathrm{e}=6.5\!\times\!10^{18} \, \mathrm{m}^{-3}$ shows both, a capacitive and inductive component and is therefore identified as E/H-hybridmode. Finally, the microwave measurements are accurately reproduced in the high power regime, showing a coupling efficiency of $\geq60\,\%$. However, with decreasing $P_{\mathrm{abs}}$ and $n_\mathrm{e}$ deviations occur, presumably due to the idealized description of the electrical network in the model. In summary, the presented model provides an accurate description of spatially resolved and global quantities. Significant deviations only occur for volume averaged parameters at lower electron densities, owing to the drastic simplifications of the global model. This is of minor importance as the MMWICP is designed to operate at $n_\mathrm{e}\geq10^{19} \, \mathrm{m}^{-3}$. Future research should address the lack of spatial resolution by a three-dimensional plasma model.

 %and distorts the loss power to unrealistically high values.
\newpage
\section{Figures}

\pagebreak
\pagebreak
\pagebreak

\pagebreak

\pagebreak

	%The phase corresponding to the calibration plane for the experimental data has been shifted. Thus the resonance curves are rotated and show only the contribution of the resonator.
\pagebreak

\pagebreak

\pagebreak

\pagebreak

\pagebreak
\section{Acknowledgements}

The authors thank J{\'o}n T{\'o}mas Gudmundsson for sharing his chemical reaction set and valuable discussions. We gratefully acknowledge the support by Deutsche Forschungsgemeinschaft DFG via SFB 1316 "Transient Atmospheric Pressure Plasmas: from plasmas to liquids to solids".

\pagebreak

\pagebreak


\newpage
\begin{references}

\bibitem{ref1} H.-E.~Porteanu, R.~Gesche, K.~Wandel, \textit{Plasma Sources Sci. Technol.} \textbf{22}, 035016 (2013)

\bibitem{ref2} I.~Stefanović, N.~Bibinov, H.-E.~Porteanu, M.~Klute, R.P.~Brinkmann, P.~Awakowicz, \textit{Plasma Sources Sci.~Technol.} \textbf{27}, 12LT01 (2018)
%"`Bright idea, radio-frequency light sources"',

\bibitem{ref3} H.-E.~Porteanu, I.~Stefanović, N.~Bibinov, M.~Klute, P.~Awakowicz, R.~P.~Brinkmann, W.~Heinrich, \textit{Plasma Sources Sci.~Technol.} \textbf{28}, 035013 (2019)
%"`Bright idea, radio-frequency light sources"',

\bibitem{application1} U.~Schulz, P.~Munzert, H.~Kaiser, \textit{Surf. Coat. Technol.} \textbf{142-144}, 507 (2001)

\bibitem{application2} E.~Tatarova, N.~Bundaleska, F.~M.~Dias, D.~Tsyganox, R.~Saavedra, C.~M.~Ferreira, \textit{Plasma Sources Sci.~Technol.} \textbf{22}, 065001 (2013)

\bibitem{application3} A.~I.~Al-Shamma'a, S.~R.~Wylie, J.~Lucas, C.~F.~Paul, \textit{J. Phys. D: Appl. Phys.} \textbf{34}, 2734 (2001)

\bibitem{application4} J.~Giersza, K.~Jankowskia, A.~Ramszab, E.~Reszkec,  \textit{Spectrochim. Acta} B \textbf{147}, 51 (2018)

\bibitem{applicationatm1} G.~Y.~Park, S.~J.~Park, M.~Y.~Choi, I.~G.~Koo, J.~H.~Byun, J.~W.~Hong, J.~Y.~Sim, G.J.~Collins, J.~K.~Lee, \textit{Plasma Sources Sci.~Technol.} \textbf{22}, 065001 (2013)


\bibitem{application5}  A.~Bogaerts, X.~Tu, J.~C.~Whitehead, G.~Centi, L.~Lefferts, O.~Guaitella, F.~Azzolina-Jury, H.-H.~Kim, A.~B.~Murphy, W.~F.~Schneider, \textit{J. Phys. D: Appl. Phys.} \textbf{53} 443001 (2020)


\bibitem{ref4} M.~Klute, H.-E.~Porteanu, I.~Stefanović, W.~Heinrich, P.~Awakowicz, R.~P.~Brinkmann, \textit{Plasma Sources Sci.~Technol.} \textbf{29}, 065018 (2020)

\bibitem{ref5} E.~G.~Thorsteinsson and J.~T.~Gudmundsson, \textit{Plasma Sources Sci.~Technol.} \textbf{18}, 045002 (2009)

\bibitem{ref7} Y.~Itikawa, \textit{Journal of Physical and Chemical Reference Data} \textbf{35}, 31 (2006)

\bibitem{Sakiyama_2012} Y.~Sakiyama, D.~B.~Graves, H.~W.~Chang, T.~Shimizu, G.~E.~Morfill, \textit{J. Phys. D: Appl. Phys.} \textbf{45} 425201 (2012)

\bibitem{globalmodel2} C.~Lee, M.~A.~Lieberman, H.~W.~Chang, T.~Shimizu, G.~E.~Morfill, \textit{J. Vac. Sci.
Technol. A} \textbf{13} 368 (1995)


\bibitem{ref10} E.~Kemaneci, E.~Carbone, M.~Jimenez-Diaz, W.~Graef, S.~Rahimi, J.~van Dijk, G.~Kroesen, \textit{J. Phys. D: Appl. Phys.} \textbf{48}, 435203 (2015)



\bibitem{ref11} E.~Kemaneci, F.~Mitschker, J.~Benedikt, D.~Eremin, P.~Awakowicz, R.~P.~Brinkmann, \textit{Plasma Sources Sci.~Technol.} \textbf{28}, 115003 (2019)

\bibitem{ref12} E.~Kemaneci, F.~Mitschker, M.~Rudolph, D.~Szeremley, D.~Eremin, P.~Awakowicz, R.~P.~Brinkmann, \textit{J. Phys. D: Appl. Phys.} \textbf{50}, 245203 (2017)

\bibitem{ref8} N.~Bibinov, D.~Dudek, P.~Awakowicz, J.~Engemann, \textit{J. Phys. D: Appl. Phys.} \textbf{40}, 7372 (2007) 

\bibitem{ref6} B.~Offerhaus, J.~W.~Lackmann, F.~Kogelheide, V.~Bracht, R.~Smith, N.~Bibinov, K.~Stapelmann and P.~Awakowicz, \textit{Plasma Process. Polym.} \textbf{14}, 1600255 (2017)

\bibitem{ref20} S.~Steves, T.~Styrnoll, F.~Mitschker, S.~Bienholz, N.~Bibinov, P.~Awakowicz, \textit{J. Phys. D: Appl. Phys.} \textbf{46}, 445201 (2013)


%	doi = {10.1088/0022-3727/45/42/425201},
%	url = {https://doi.org/10.1088%2F0022-3727%2F45%2F42%2F425201},
%	year = 2012,
%	month = {oct},
%	publisher = {{IOP} Publishing},
%	volume = {45},
%	number = {42},
%	pages = {425201},
%	author = {Yukinori Sakiyama and David B Graves and Hung-Wen Chang and Tetsuji Shimizu and Gregor E Morfill},
%	title = {Plasma chemistry model of surface microdischarge in humid air and dynamics of reactive neutral species},
%	journal = {Journal of Physics D: Applied Physics},





\end{references}
\end{document}